\documentstyle[aps,pra]{revtex}

\begin{document}
\twocolumn[\hsize\textwidth\columnwidth\hsize\csname
@twocolumnfalse\endcsname

\draft

\title{Exact solutions of $n$-level systems and gauge theories}

\author{Merced Montesinos$^{a,b,c}$\footnote{e-mail: merced@fis.cinvestav.mx},
Abdel P\'erez-Lorenzana$^{b,d}$\thanks{e-mail: 
aplorenz@Glue.umd.edu}}

\address{
$^a$ Centre de Physique Th\'eorique, CNRS Luminy, F-13288 Marseille, France.}
\address{
$^b$Departamento de F\'{\i}sica, Centro de Investigaci\'on y de 
Estudios Avanzados del I.P.N.,\\
Av. I.P.N. No. 2508, 07000 Ciudad de M\'exico, M\'exico.}
\address{
$^c$ Department of Physics and Astronomy,
University of Pittsburgh, Pittsburgh, PA 15260, USA.}
\address{
$^d$Department of Physics, University of Maryland, College Park, Maryland 
20742, USA.}

\date{\today}

\maketitle

\begin{abstract}
We find a relationship between unitary transformations of the dynamics of 
quantum systems with time-dependent Hamiltonians and gauge theories. In 
particular, we show that the nonrelativistic dynamics of spin-$\frac12$ 
particles in a magnetic field $B^i (t)$ can be formulated in a natural 
way as an $SU(2)$ gauge theory, with the magnetic field $B^i(t)$ playing the 
role of the gauge potential $A^i$. The present approach can also be 
applied to systems of $n$ levels with time-dependent potentials, $U(n)$ 
being the gauge group. This geometric interpretation provides a powerful 
method to find exact solutions of the Schr\"odinger equation. The root of 
the present approach rests in the Hermiticity property of the Hamiltonian 
operators involved. In addition, the relationship with true gauge 
symmetries of $n$-level quantum systems is discussed. 
\end{abstract}


\vskip10pt]

The discovery of the quantum nature of matter fields is one of the 
most important events in physics in this century. It has been realized 
that the quantum nature of matter rests on a more fundamental level than 
the corresponding classical limit. The last one emerges as a suitable 
limit case of quantum behavior \cite{Dirac}. Also, gauge as well as 
geometric aspects of quantum systems are one of the most striking 
results in quantum theory \cite{Ber84,Aha87,Sam88,Rov98}. Gauge theories 
appear in many branches of physics \cite{Book}. They are the natural 
language to incorporate internal symmetries of systems. 

On the other hand, the physics of systems of two levels has been one of the 
most interesting research lines followed up to now. In particular, the 
dynamics of spin-$\frac12$ particles in time-dependent magnetic fields is a 
common physical situation in neutron interferometry 
\cite{Badurek,Wagh}. This type of physical 
application belongs to $n$-level systems with time-dependent potentials in 
quantum mechanics. Nevertheless, to find exact solutions of the Schr\"odinger 
equation with time-dependent potentials for $n$-level systems is a hard 
problem. The standard theoretical techniques employed to solve this type 
of physical situation are limited. 

Here, we combine the topics of exact solutions and gauge theories. The 
final result is twofold. First, we exhibit the geometric structure underlying 
unitary transformations associated with $n$-level quantum systems 
characterized by time-dependent Hamiltonians. Second, this formalism 
provides a powerful method to find exact solutions to a very general 
class of time-dependent Hamiltonians. Moreover, because the present formalism 
is {\it exact}, it is a serious alternative for the standard techniques 
allowed in the literature to study this type of physical situation, namely, 
unitary evolution and time-dependent perturbation theory.

Now, let us go to the nonrelativistic dynamics of spin-$\frac12$ 
particles (we mean the spinorial part of the Hamiltonian of 
spin-$\frac12$ particles). The nonrelativistic dynamics of 
spin-$\frac12$ particles in a magnetic field 
${\vec B} (t) = (B^1 (t) , B^2 (t) , B^3 (t))$ is given by the 
Schr\"odinger-Pauli
equation
\begin{eqnarray}
{\widehat H } (t) |\psi (t) \rangle =
-\frac{1}{\hbar} g_s \mu_B {\vec B} (t) \cdot 
{\vec S}\,\, |\psi (t) \rangle
=i \hbar \frac{d}{dt} |\psi (t) \rangle\, , \label{Pauli}
\end{eqnarray}
where $\mu_B =\frac{e\hbar}{2 m_e c}$ is the Bohr magneton, 
$S_i =\frac{\hbar}{2} \sigma_i$ with $\sigma_i$ the Pauli 
matrices, and $g_s$ depends on the specific spinor field, for instance, 
$g_s=2$ for electrons. It is clear that Eq. (\ref{Pauli}) can be expressed as 
\begin{eqnarray}
{\nabla} |\psi (t) \rangle := 
\left ( \frac{d}{dt} +  A^i (t) J_i \right )  
\pmatrix {\psi_1 (t) \cr \psi_2 (t)} =0 \, ,\label{connection}
\end{eqnarray}
with $A^i (t) = \frac{1}{\hbar} g_s \mu_B B^i (t)=:\mu B^i$. The matrix 
composed of the magnetic field in Eq. (\ref{connection}) is valued in the 
Lie algebra of the $SU(2)$ group, $su(2)$, namely, 
$A (t) = A^i (t) J_i$, with $J_i =-\frac{i}{2} \sigma_i$ the 
(traceless skew-Hermitian $2\times 2$) matrices which satisfy 
$[J_i , J_j ]=\epsilon_{ij}\,^k J_k$ where $\epsilon_{ij}\,^k$ are the 
structure constants (Levi-Civit\`a symbol)   
\begin{eqnarray}
A (t) & = & -\frac{\mu}{2} 
\left (
\begin{array}{lr}
i B^3 (t) & i B^1 (t) + B^2 (t) \\
i B^1 (t) - B^2 (t) & - i B^3 (t)
\end{array}
\right ) \, .\label{YM}
\end{eqnarray}
This matrix has a geometric interpretation, in the present formalism, as 
we will see in a moment. Here, we are interested in finding exact solutions 
of Eq. (\ref{connection}). Let us assume that 
we have an exact solution of the Schr\"odinger-Pauli equation, i.e., that 
we have analytical expressions for the spinor field 
\begin{eqnarray}
|\psi (t) \rangle =\pmatrix {\psi_1 (t) \cr \psi_2 (t)}
\end{eqnarray}
as well as for 
the magnetic field ${\vec B} (t)$ in such a way that when we put 
the pair (${\vec B}$, $|\psi (t) \rangle$) in Eq. (\ref{connection}), the 
equation holds. Next, we make the transformation 
\begin{eqnarray}
|\psi' (t) \rangle =
\pmatrix {\psi'_1 (t) \cr \psi'_2 (t)}= G(t)
\pmatrix {\psi_1 (t) \cr \psi_2 (t)} = G(t) 
|\psi (t) \rangle  \label{transformation}
\end{eqnarray}
on the spinor field $|\psi (t) \rangle$, where the matrix $G(t)$ is 
in $SU(2)$. Is the spinor 
$|\psi' (t) \rangle$ a solution of the Schr\"odinger-Pauli 
equation? If so, what is the new magnetic field 
${\vec B}' (t)$? The answer is in the affirmative. By plugging 
Eq. (\ref{transformation}) into 
Eq. (\ref{connection}), we find 
\begin{eqnarray}
{\nabla}' |\psi' (t) \rangle & := &  
\left ( \frac{d}{dt} +  {A'}^i (t) J_i \right ) |\psi' (t) \rangle \nonumber\\
& = & G (t) \left ( \frac{d}{dt} +  {A}^i (t) J_i \right ) 
|\psi (t) \rangle \nonumber\\
& = & G(t) {\nabla} |\psi (t) \rangle = 0\, ,\label{NNN}
\end{eqnarray}
provided that the matrix composed of the magnetic field (\ref{YM}) 
transforms as
\begin{eqnarray}
A' (t) = G(t) A(t) G^{-1} (t) +G (t) \frac{d}{dt} G^{-1} (t)\, ,
\label{YMtrans}
\end{eqnarray}
which is precisely the transformation rule for a Yang-Mills 
connection, $SU(2)$ being the local (in time) gauge group. Thus, it is clear 
that if the pair $( A^i (t), |\psi (t) \rangle)$ is a solution 
of Eq. (\ref{connection}), then the pair
$( {A'}^i (t), |\psi' (t) \rangle)$ {\it is} also a new solution of the 
Schr\"odinger-Pauli equation. From Eq. (\ref{YMtrans}) it is easy to compute 
the new magnetic field ${\vec B}' (t)$. If we parametrize the $G(t)$ matrix 
as
\begin{eqnarray}
G (t) & = & e^{\vec{\alpha} (t) \cdot {\vec J}  }  \nonumber\\
& = & \cos{\left (\frac{\alpha (t)}{2} \right )}- 
i {\widehat \alpha} (t) \cdot 
{\vec \sigma} \sin{\left ( \frac{\alpha (t)}{2} \right )}\, , \label{groupm}
\end{eqnarray} 
and we insert this expression into Eq. (\ref{YMtrans}), a straightforward 
calculation yields
\begin{eqnarray}
{\vec B}' (t) & = & {\vec B} (t) \cos{\alpha}(t) - ({\vec B} (t) \times
\widehat{\alpha} (t) ) \sin{\alpha} (t) + \nonumber\\
& & + 2 \widehat{\alpha} (t) ({\vec B} (t) \cdot
\widehat{\alpha} (t) ) \sin^2{ \left ( \frac{\alpha (t)}{2}\right )} 
+ {1\over\mu}{d\over dt}\vec{\alpha} (t) \, , \label{field}
\end{eqnarray}
with $\vec{\alpha} (t) = ( \alpha^1 (t), \alpha^2 (t),\alpha^3 (t))$, 
$\alpha (t)=|\vec{\alpha} (t)|$, and 
${\widehat \alpha} (t)= \frac{\vec{\alpha} (t)}{\alpha (t)}$. The new 
magnetic field $\vec{B}' (t)$ has i) a new temporal dependence 
through $\cos{\alpha (t)}$ along the original direction of the
magnetic field $\vec{B} (t)$, ii) a rotation 
term around $\widehat {\alpha} (t)$, iii) a contribution
along $\widehat{\alpha} (t)$ from the projection of $\vec{B} (t)$, and
iv) a tangential contribution (last term on the right-hand side of 
Eq. (\ref{field})). Notice how general the expression for the magnetic 
field ${\vec B}' (t)$ is. It depends on three arbitrary functions 
$\alpha_1 (t)$, $\alpha_2 (t)$, and $\alpha_3 (t)$ once the magnetic 
field ${\vec B} (t)$ has been given. Actually, we can consider the 
three components of ${\vec B} (t) = ( B^1 (t), B^2 (t), B^3 (t))$ also as 
arbitrary functions because in the formalism ${\vec B} (t)$ is free and it 
has to be specified.

Let us summarize the results so far. The dynamics of 
spin-$\frac12$ particles in a
magnetic field $\vec{B} (t)$ can be interpreted as the compatibility 
condition of the Yang-Mills covariant derivative on the spinor field,
namely, ${\nabla} |\psi (t)\rangle =0$ with
$A^i (t) = \mu B^i (t)$ the gauge potential of the
$SU(2)$ gauge theory (see Eq. (\ref{connection})). Under the 
transformation (\ref{transformation}), the 
new spinor field $|\psi' (t) \rangle$ and the new magnetic field 
$\vec{B} ' (t)$ (given by Eq. (\ref{field})) also do 
satisfy the Schr\"odinger-Pauli equation 
(\ref{connection}) \cite{Note}. Readers interested in applying the 
present formalism have to use the general expressions 
(\ref{transformation}), (\ref{groupm}), and (\ref{field}). Now, we give 
an example.

{\it Example.} From constant to rotating magnetic fields. Let us 
take $\vec{B}= B_0\hat k$, with $B_0$ a constant. The spinor 
$|\psi (t) \rangle$ which is a solution of the Schr\"odinger-Pauli equation 
with $\vec{B}= B_0\hat k$ is
\begin{eqnarray}
|\psi (t)\rangle & = & 
\left ( 
\begin{array}{l}
 \psi_1 (t) \\
 \psi_2 (t)
\end{array}
\right ) =
\left ( 
\begin{array}{l}
e^{-i \left ( \frac{\mu B_0 }{2} t \right ) } \psi_{10} \\
e^{+i \left ( \frac{\mu B_0 }{2} t \right ) } \psi_{20}
\end{array}
\right )\, .
\end{eqnarray}
where $\psi_{10}$, $\psi_{20}$ are two complex constants that fix the 
orientation of the original state. We are ready to find the 
{\it new} magnetic field ${\vec B}' (t)$ and 
the {\it new} spinor field $|\psi' (t)\rangle$. We give 
$\vec{\alpha} = \alpha(t)\hat\i$ with $\alpha (t)$ an arbitrary function of 
time. Thus, from Eq. (\ref{field}) we have the expression of the new magnetic
field
\begin{eqnarray} 
{\vec B}' (t) & = & B_0 \left [ \cos{\alpha(t)} \hat{k} - \sin{\alpha (t)}
\hat{\j} \right ] + 
\frac{1}{\mu} {\dot \alpha} (t) \hat{\i} \, ,
\end{eqnarray}
where ${\dot \alpha} (t)= \frac{d \alpha (t)}{dt}$. ${ \vec B}' (t)$ is a 
rotating magnetic field around the $x$ axis. This fact becomes clear if we 
set for simplicity $\alpha(t) = \omega t$, to get
\begin{eqnarray}
{\vec B}' (t) & = & B_0 \left [ \cos{(\omega t)} \hat{k} - 
\sin{(\omega t)} \hat{\j} \right ] + \frac{\omega}{\mu} \hat{\i} .
\end{eqnarray}
By inserting $\vec{\alpha} = \alpha(t)\hat\i$ into Eq. (\ref{groupm}) we 
obtain the explicit form of the group element 
$G(t)=\cos(\alpha/2) - i\sigma_1 \sin(\alpha/2)$, and the expression for the 
new spinor field is 
\begin{eqnarray}
|\psi' (t) \rangle & = & G(t) |\psi (t) \rangle = 
\left ( 
\begin{array}{l}
 {\psi'}_1 (t) \\
 {\psi'}_2 (t)
\end{array}
\right ) =\nonumber
\end{eqnarray}
\begin{eqnarray}
& = &
\left(
\begin{array}{l}
\cos(\frac{\alpha}{2}) e^{-i \left ( \frac{\mu B_0}{2}\right ) t}
\psi_{10} -  i \sin(\frac{\alpha}{2}) 
e^{+ i\left (\frac{\mu B_0}{2} \right ) t} 
\psi_{20} \\
\cos(\frac{\alpha}{2} ) e^{+i \left (\frac{\mu B_0}{2} \right ) t}
\psi_{20} - i \sin(\frac{\alpha}{2}) 
e^{- i\left (\frac{\mu B_0}{2} \right ) t}
\psi_{10} 
\end{array}
\right) \, .
\end{eqnarray} 

Now, we come back to the geometric structure of spin-$\frac12$ 
dynamics. As in every gauge theory, we have also the 
{\it covariant derivative} on the adjoint bundle \cite{Book}. In the case of
spin-$\frac12$ particles in a magnetic field, the equation of motion 
which resides in the associated bundle is
\begin{eqnarray}
{\nabla} n^i := \frac{d n^i (t)} {dt} + 
\epsilon^i\,_{jk} A^j (t) n^k (t) =0 \, ,
\label{adequation}
\end{eqnarray}
with $n^i (t) = \langle \psi (t)| \sigma^i | \psi (t) \rangle$. Notice that
$n (t)=n^i (t) dt \otimes R_i$ is a local (in time) one-form valued in the 
adjoint representation of $su(2)$, with $R_i$ $3\times 3$ 
matrices satisfying $[R_i , R_j ]=\epsilon_{ij}\,^k R_k$. Under the 
transformations (\ref{transformation}), (\ref{groupm}), and 
(\ref{field}), we get
\begin{eqnarray}
\frac{d {n'}^i (t)}{dt} + \epsilon^i\,_{jk} {A'}^j (t) {n'}^k (t) = 0 \, ,
\end{eqnarray}
with ${n'}^i (t) = \langle \psi' (t)| \sigma^i | \psi' (t) \rangle$, 
${A'}^i (t) = \mu {B'}^i (t)$ as was expected. Notice that 
Eq. (\ref{adequation}) follows from application of the Ehrenfest theorem.  

Now, some comments on the {\it direct problem} and 
{\it inverse technique} approaches in quantum mechanics. The direct problem 
consists in giving the magnetic field $\vec{ B} (t)$ and finding the 
spinor field $|\psi (t)\rangle$ which satisfies the Schr\"odinger-Pauli 
equation. On the other hand, inverse techniques try to find the
 possible magnetic fields ${\vec B} (t)$ that are compatible with a 
given spinor field $|\psi (t)\rangle$. The present formalism 
is something in between these two approaches. The proposal developed here 
determines simultaneously {\it both} the magnetic field ${\vec B}' (t)$ and 
the spinor field $|\psi' (t)\rangle$ from a given initial
pair $({\vec B} (t)\, ,\, |\psi (t)\rangle )$, and the new pair 
$( {\vec B}' (t) , |\psi' (t)\rangle)$ satisfies also the 
Schr\"odinger-Pauli equation. 

Also, let us emphasize that the present formalism is not that of
rotating-frame techniques \cite{WaghII}, for the physical interpretation 
underlying both cases is completely different. Here, we interpret 
the {\it old} and {\it new} pairs $({\vec B} (t)\, ,\, |\psi (t)\rangle )$ as 
{\it different} physical solutions associated with the dynamics of 
spin-$\frac12$ particles. In \cite{WaghII}, as opposed to 
here, the rotating-frame technique is only a tool to understand the 
{\it same} physical situation from the point of view of 
the laboratory frame as well as from 
the rotating frame viewpoint. Even though these results would admit an 
interpretation in the rotating-frame formalism, that point of view is 
restricted to two-level systems and it does not apply to the generic case of 
$n$-level systems that we consider from now on. 

The generalization to systems with time-dependent potentials is 
straightforward.  This type of physical situation appears, for instance, when
systems are perturbed with time-dependent potentials. Without loss of 
generality, let us focus in the nondegenerate case. The Schr\"odinger 
equation of a $n$-level system with time-dependent potential is
\begin{eqnarray}
{\widehat H} (t) |\psi (t) \rangle = 
i\hbar \frac{d}{dt} |\psi (t) \rangle\, .\label{time}
\end{eqnarray}  
Here, ${\widehat H} (t)$ is a time-dependent $n\times n$ matrix, and 
$|\psi (t) \rangle$ is an $n$-dimensional vector. We are interested in 
finding exact solutions to Eq. (\ref{time}). Because
${\widehat H} (t)$ is a Hamiltonian, ${\widehat H} (t)$ is a $n\times n$ 
Hermitian matrix. This means that the term 
$\frac{i}{\hbar} {\widehat H} (t)$ in
\begin{eqnarray}
\frac{d}{dt} |\psi (t) \rangle + \frac{i}{\hbar}
{\widehat H} (t) |\psi (t) \rangle = 0 \label{newtime}
\end{eqnarray} 
is a $n\times n$ {\it skew-hermitian} matrix, i.e., 
$\frac{i}{\hbar} {\widehat H} (t)$ is valued in the (real) Lie 
algebra of the $U(n)$ group, $u(n)$, which means we can express 
$\frac{i}{\hbar} {\widehat H} (t)$ as
\begin{eqnarray}
\frac{i}{\hbar} {\widehat H} (t) = A^i (t) J_i \, ,
\end{eqnarray}
where $J_i$, $i=1,...,n^2$, are the generators of $u(n)$. Notice that 
we are working in the fundamental representation of the $U(n)$ 
group ($|\psi (t) \rangle$ is an $n$-component 
vector and $J_i$ are $n \times n$ skew-hermitian matrices). This 
is the key property we are going to exploit. Therefore, we can 
rewrite the Schr\"odinger equation (\ref{newtime}) as
\begin{eqnarray}
{\nabla} |\psi (t) \rangle := \left ( \frac{d}{dt} +  
A^i (t) J_i \right ) |\psi (t) \rangle = 0\, . \label{Mills} 
\end{eqnarray}
The geometric meaning of the equation of motion is very clear. Dynamics 
reduces to the compatibility condition of the Yang-Mills covariant 
derivative on the state vector. This geometric meaning is very 
suggestive. It induces us to make make the transformation 
\begin{eqnarray}
|\psi' (t) \rangle = G(t) |\psi (t) \rangle  \label{TTT}
\end{eqnarray} 
on the state vector $|\psi (t) \rangle$, with $G(t)$ an $U(n)$ matrix. By
doing this, we get
\begin{eqnarray}
{\nabla}' |\psi' (t) \rangle & := & \left ( \frac{d}{dt} +  
{A'}^i (t) J_i \right ) |\psi' (t) \rangle \nonumber\\
& = & G (t) \left ( \frac{d}{dt} +  A^i (t) J_i \right ) 
|\psi (t) \rangle = 0\, ,
\end{eqnarray}
with
\begin{eqnarray}
{A'}^i (t) J_i = G(t) A^i (t) J_i G^{-1} (t) + 
G (t) \frac{d}{dt} G^{-1} (t)\, ,
\end{eqnarray}
the transformation law for a Yang-Mills connection valued in 
$u(n)$. A beautiful result comes from 
these last three expressions. If the pair ($A^i (t)$, $|\psi (t) \rangle$) 
is a solution of the equation of motion (\ref{Mills}), i.e., if the 
analytical expressions of $A^i (t)$,  and $|\psi (t) \rangle$ 
satisfy Eq. (\ref{Mills}), then the {\it new} pair 
(${A'}^i (t)$, $|\psi' (t) \rangle$) {\it is} also a solution of 
Eq. (\ref{Mills}). Here, we have written our equations in terms of 
the concepts and notions of Yang-Mills theories \cite{Book}. This was done 
to show the geometric structure which underlies in the present 
formalism. Readers interested in the expression of the {\it new} 
Hamiltonians ${\widehat H}' (t)$ have to consider 
\begin{eqnarray}
{\widehat H}' (t) = G(t) {\widehat H} (t) G^{-1} (t) + 
\frac{\hbar}{i} G(t) \frac{d}{dt} G^{-1} (t)\, , \label{HHH}
\end{eqnarray}  
as well as the transformation of the state vector (\ref{TTT}). Notice that 
there are $n^2$ time-dependent functions associated with the new 
Hamiltonian and state vector (see Eqs. (\ref{TTT}) and (\ref{HHH})) because 
the gauge group involved is $U(n)$.

As in the case of spin-$\frac12$ systems, the {\it covariant derivative} 
in Eq. (\ref{Mills}),  
\begin{eqnarray} 
\nabla:=  \frac{d}{dt} +  A^i (t) J_i 
\end{eqnarray}
is nothing but the covariant derivative in the associated vector bundle 
$E=P \times C^n$, corresponding to the principal bundle $P(R^+ , G)$; 
$R^+$ means the time evolution of the system and $C$ the complex 
numbers of each one of the entries in $|\psi (t)\rangle$ . As was
mentioned, the representation of the matrices of the Lie algebra in the 
covariant derivative belongs to the same dimension of the state 
$|\psi \rangle$. The Yang-Mills derivative also allows us to work with 
differentrepresentations of the Lie algebra (a fact that is not used here). 

Note also that (as in every gauge theory) we have another 
associated vector bundle, the adjoint vector 
bundle $E_g = P \times _{Ad} g$, where the covariant 
derivative is given by
\begin{eqnarray}
{\nabla} \lambda^i (t) := \frac{d}{dt}\lambda^i (t) + 
f_{jk}\,^i A^j (t) \lambda^k (t) \, ,
\end{eqnarray}
where $\lambda$ is a local (in time) one-form valued in the adjoint 
representation of $u(n)$, 
$\lambda =\lambda^i (t) J_i \otimes dt$. Therefore, the adjoint 
equation of motion of the Schr\"odinger equation exists and is given by 
\begin{eqnarray}
{\nabla} \lambda^i (t) = \frac{d}{dt}\lambda^i (t) + 
f_{jk}\,^i A^j (t) \lambda^k (t) = 0\, , \label{adjointeq}
\end{eqnarray}
which follows from application of the Ehrenfest theorem, 
$\lambda^i (t) = \langle \psi (t) | i J_i | \psi (t) \rangle$. Under the 
transformation (\ref{TTT}) the adjoint equation transforms as
\begin{eqnarray}
{\nabla}' {\lambda '}^i (t) = \frac{d}{dt} {\lambda '}^i (t) + 
f_{jk}\,^i {A'}^j (t) {\lambda'}^k (t) = 0\, ,
\end{eqnarray} 
with ${\lambda '}^i (t) = \langle {\psi'} (t) | i J_i | {\psi'} (t) \rangle$. 

Let us summarize our results. We have found that there is 
a natural geometric structure between exact solutions of 
time-dependent potentials in $n$-level quantum systems and gauge 
theories. It has to be clear that in the standard application 
of gauge theories, there is a fixed 
system where the gauge symmetry is a property of the {\it same} 
system. Here, on the contrary, we use gauge theories as the geometric 
structure that allows us to relate {\it different} quantum systems 
(in the sense of different solutions to the equation of motion). At the 
same time, this geometric interpretation provides an alternative 
method for the traditional (and limited) ones allowed in the literature 
to attack physical situations with time-dependent Hamiltonians. 

Finally, the possibility exists of interpreting 
the present formalism as a {\it true} gauge theory. This is 
as follows. Let us fix a {\it specific} $n$-level system 
from the very beginning. Next, we ask for those allowed transformations 
${\widetilde G} (t)$ on the state vector $|\psi (t) \rangle$,
\begin{eqnarray}
|{\widetilde \psi} (t) \rangle = {\widetilde G} (t) |\psi (t) \rangle \, , 
\end{eqnarray}   
in such a way that the new Hamiltonian {\it is} exactly the same as 
the old one; i.e., 
\begin{eqnarray}
{\widehat H} (t) = {\widetilde G} (t) {\widehat H} (t) 
{\widetilde G}^{-1} (t) + \frac{\hbar}{i} {\widetilde G } (t) 
\frac{d}{dt} {\widetilde G}^{-1} (t)  
\end{eqnarray} 
is satisfied. We call this case the {\it strong} gauge condition 
of the system. More precisely, this transformation does give us all the 
state vectors $|\psi (t) \rangle$ belonging to the same equivalence 
class defined by a specific gauge potential $A^i (t)$. From this 
viewpoint, ${\widetilde G} (t)$ is a {\it non-Abelian} phase that relates 
the family of state vectors 
$\{ |\psi (t)\rangle \}$ associated with the {\it same} 
Hamiltonian. One can think of 
this condition as a very restrictive one. There is, in fact, also a 
{\it weak} condition. This last condition takes advantage of the adjoint 
equation of motion (\ref{adjointeq}). Instead of looking at the 
equivalence class of the state vectors 
$|\psi (t) \rangle \sim |{\widetilde \psi} (t) \rangle$ defined 
by the same Hamiltonian, we look at the equality of the mean values
$\lambda^i (t) = {\widetilde\lambda}^i (t)$, computed with the state 
vectors $|\psi (t) \rangle $ and 
$|{\widetilde\psi} (t) \rangle = {\widetilde G}(t) 
|\psi (t) \rangle$, respectively. Therefore, from the adjoint equation 
(\ref{adjointeq}), in this case it follows that 
$f_{jk}\,^i ( A^j (t) -{A'}^j (t) ) \lambda^k (t)=0$. So the 
transformation ${A'} (t) = A^i (t) + \Omega^i (t) $ with $\Omega^i (t)$ 
in the same direction as $\lambda^i (t)$, leads to the same physics (same 
analytical expressions for the mean values). An interesting application is 
precisely the dynamics of spin-$\frac12$ particles in magnetic fields. In 
this case, the {\it strong} gauge condition gives all the state vectors 
compatible with a certain magnetic field $B^i (t)$, while the {\it weak} 
gauge condition $\epsilon_{jk}\,^i ({B'}^j (t) - B^j (t)) n^k (t) =0$ 
gives the same analytical expressions for the mean values of 
$n^i (t)=\langle \psi (t) |\sigma^i |\psi (t)\rangle$, 
associated with $B^i (t)$, as well as for 
${\widetilde n}^i (t)=\langle {\widetilde \psi} (t) |\sigma^i | 
{\widetilde \psi} (t)\rangle $, associated 
with ${B'}^j (t) =B^j (t) + \Omega^j (t)$, where $\Omega^j (t)$ is a 
time-dependent three-dimensional vector in the same direction of 
$n^j (t)$. Finally, coming back to the generic case, the implications of the 
weak gauge condition on the cyclic evolution of systems in the {\it space of 
mean values} $\lambda^i (t)$ and the relationship of this cyclic evolution 
with geometric phases in the space of parameters $A^i$ are 
straightforward. In the case of a spin-$\frac12$ system, closed loops, in the 
space of the mean values $n^i$ of the components of the spin, relate the 
geometric phases associated with $B^i(t)$ and with $B^i (t) + 
\Omega^i (t)$, respectively, when $B^i (t)$ and $\Omega^i (t)$ are periodic 
functions.

\section*{Acknowledgments}
Thanks are due to Roberto De Pietri and Carlo Rovelli for comments. Also, we 
thank D. J. Fern\'andez C. and O. Rosas-Ortiz for carefully 
reading the manuscript and for their valuable comments. A.P.L. thanks all the 
members of the Department of Physics of the University of Maryland for 
their kind hospitality. Also, the authors acknowledge financial support 
provided by the {\it Sistema Nacional de Investigadores} of the 
Secretar\'{\i}a de Educaci\'on P\'ublica (SEP) and by CONACyT of Mexico. 


\end{document}